# Trustworthy V2G scheduling and energy trading: A blockchain-based framework

Yunwang Chen[a], Xiang Lei[a], Songyan Niu[a,b] and Linni Jian[a,b,*]

a Department of Electronic and Electrical Engineering, Southern University of Science and Technology, Shenzhen, 518055, China
b Southern University of Science and Technology Jiaxing Research Institute, Jiaxing, 314000, China
* Correspondence: jianln@sustech.edu.cn; Tel.: +86-755-8801-8525

**Abstract:** The rapid growth of electric vehicles (EVs) and the deployment of vehicle-to-grid (V2G) technology pose significant challenges for distributed power grids, particularly in fostering trust and ensuring effective coordination among stakeholders. Establishing a trustworthy V2G operation environment is crucial for enabling large-scale EV user participation and realizing V2G's potential in real-world applications. In this paper, an integrated scheduling and trading framework is developed to conduct transparent and efficacious coordination in V2G operations. In blockchain implementation, a cyber-physical blockchain architecture is proposed to enhance transaction efficiency and scalability by leveraging smart charging points (SCPs) for rapid transaction validation through a fast-path practical byzantine fault tolerance (fast-path PBFT) consensus mechanism. From the energy dispatching perspective, a game-theoretical pricing strategy is employed and smart contracts are utilized for autonomous decision-making between EVs and operators, aiming to optimize the trading process and maximize economic benefits. Numerical evaluation of blockchain consensus shows the effect of the fast-path PBFT consensus in improving systems scalability with a balanced trade-off in robustness. A case study, utilizing real-world data from the Southern University of Science and Technology (SUSTech), demonstrates significant reductions in EV charging costs and the framework's potential to support auxiliary grid services.

## 1. Introduction

Amidst the growing global preoccupation with energy sustainability and environmental considerations, a resounding consensus has emerged among nations towards the objectives of transportation electrification [1]. This focus has gained further prominence due to the looming climate crisis and the pressing need to curtail fossil fuel consumption. Notably, even in the wake of a pandemic-induced economic downturn, the resilience of the electric vehicle (EV) market has been strikingly evident, with EV registrations surging by a remarkable 41% in the year 2020 [2]. This robust growth not only underscores the increasing popularity of EVs but also presents the enduring global commitment towards sustainable transportation alternatives. It is predicted that more than 250 million electric vehicles will be on roadways by the year 2030 [3]. As this transition to electrified transportation gains momentum, it becomes increasingly imperative to address the various challenges and opportunities that accompany this paradigm shift comprehensively. The intrinsic high-power characteristics of EVs accentuate their role as substantial energy consumers, thereby contributing to a notable upsurge in energy demand within the power system. If not properly managed, this considerable and dynamic charging demand could cause grid congestion, amplify demand peaks and erode power quality through voltage as well as frequency fluctuations, thereby imposing challenges on the stability and security of the power system. Meanwhile, extensive research has substantiated that EVs, beyond their transportation function, possess the potential to function as distributed energy storage systems for the power grid through vehicle-to-grid (V2G) technology [4].

The V2G technology aims to establish a bidirectional bridge connecting EVs with the power system. This approach enables EVs not only to draw power from the grid but also to contribute energy from their onboard batteries back to the grid as needed, transcending the traditional role of EVs solely as consumers of electricity into prosumers of electricity [5]. Moreover, smart V2G charging can facilitate dynamic EV charging and load-shifting grid services, thereby enhancing the overall efficiency of energy utilization. Forecasts suggest that by as early as 2030, the energy storage capacity of electric vehicle batteries alone could satisfy short-term grid storage demands, highlighting the potential of EVs to play a pivotal role in grid stability and energy management [6]. The harnessing of EV batteries carries the promise of augmenting supply flexibility, while concurrently diminishing capital expenditures and mitigating material-related emissions associated with the deployment of additional storage and power-electronic infrastructure [7]. Achieving these functionalities involves active EV engagement and coordination of EVs' charging and discharging decisions in alignment with specific objectives, such as reducing the overall system variance [4], optimizing total charging/operation cost [8], or enhancing grid stability during unforeseen demand fluctuations [9].

In the realization of V2G operations, smart charging points (SCPs) serve as crucial cyber-physical entities in facilitating the interactions between EVs and the power grid [10]. The intelligent charging point is equipped with multiple ports for communication and control, allowing it to manage the charging and discharging activities of multiple EVs. Furthermore, SCPs can establish a dynamic network with both EVs and the power grid, enabling real-time data exchange, command transmission and information storage [11]. The inherent cyber-physical nature of SCPs underscores their capacity to integrate real-time information from both the physical charging infrastructure and the digital communication network [12]. This synergistic amalgamation of physical and cyber components allows for precise monitoring and control of charging operations, rendering the charging process responsive to grid dynamics. Although the individual behavior and usage patterns of each EV are influenced by diverse uncertain factors such as driver preferences, charging routines and travel routes [13], when brought together in collective aggregation with smart charging points, the amassed EV fleet emerges as a dependable and consistent resource for grid interactions [14].

The aggregation of EVs lays the foundation for the emergence of aggregators as system operators of EVs, primarily characterized as electric vehicle aggregators (EVAs) and distribution system operators (DSOs) [15]. Both EVAs and DSOs can harness the aggregated capacity of EVs to enhance grid operations and participate in energy markets to obtain additional revenues. Acting as virtual intermediaries, EVAs leverage the combined potential of EVs to provide grid services, optimizing charging and discharging schedules to partake in electricity markets for financial gain. On the other hand, DSOs, responsible for managing the distribution grid, work to ensure grid reliability while accommodating the growing EV demand and making profits [16]. For owners of electric vehicles, engagement with an aggregator yields substantial financial incentives. These incentives may encompass compensation for contributing energy services and reduced charging costs during off-peak hours [17]. However, a prominent challenge arises in the guise of customer mistrust toward aggregators, particularly concerning the credibility of charging/discharging data and the integrity of information relayed from the aggregator to consumers. Furthermore, the presence of malicious actors within the energy trading market introduces a significant and concerning threat to both the aggregator and EVs, which encompasses many strategies such as unauthorized access, data manipulation and malicious sharing of sensitive information. To tackle these challenges, the implementation of blockchain technology has emerged as a viable solution to establish a secure and trustworthy linkage between aggregators and EV owners [18].

The blockchain serves as a distributed and immutable digital ledger, recording transactions in a secure, transparent, and tamper-proof manner that guarantees data privacy, authenticity, and integrity within the network. For aggregators, this implies that data concerning EV charging, discharging, grid interactions, and other operational specifics can be securely logged and accessed by relevant stakeholders, mitigating risks of unauthorized modifications or breaches. For EV owners, blockchain introduces the potential for real-time monitoring of EV energy contributions to the grid, ensuring precise compensation for energy provision and grid services, thereby fostering active engagement in V2G activities. Moreover, the integration of advanced cryptographic techniques within the blockchain framework ensures that personal data remains confidential and visible only to authorized entities, effectively safeguarding user privacy and building trust in V2G systems [19]. At the core of every blockchain system lies the consensus protocol, a stringent procedure guiding system nodes in reaching agreement and updating a unified, decentralized system state. However, the renowned proof-of-work (PoW) consensus protocol concerns a large waste of energy and a considerable demand for computation resources as well as the slow block validation process, while still facing vulnerabilities if a malicious entity accumulates a majority of the computation resources. These shortcomings render PoW protocol unsuitable for accommodating real-time, resource-constrained energy trading between EVA and EV owners. Meanwhile, alternative protocols, including proof-of-stake (PoS) and proof-of-authority (PoA), while aiming to ensure validation speed, cannot claim complete immunity to potential vulnerabilities [20]. The integration of EVs into distributed power grids introduces significant challenges, particularly in establishing a secure, scalable and trustworthy mechanism for energy scheduling and trading.

The interplay between cyber networks and physical energy grids positions SCPs as critical infrastructures for V2G operations. However, the lack of a transparent and fair mechanism for pricing, clearing, and settlement in V2G transactions has been a significant barrier to realizing the full potential of V2G. Additionally, ensuring that stakeholders, particularly EV owners, benefit from these systems is essential for large-scale adoption. This highlights the need for solutions that address trust, transparency, and fairness in V2G energy scheduling and trading processes. Motivated by these challenges, this study proposes a blockchain-enabled framework that utilizes smart contracts and a game-theoretical pricing strategy to establish trust and secure transparent, automated energy trading. The key focus is addressing the trust issues surrounding pricing, clearing, and settlement in V2G operations, particularly in decentralized settings where EV users and aggregators may have conflicting objectives. By employing a Stackelberg game model, a pricing strategy is provided that ensures fair economic incentives between EVA and EV while promoting transparent and verifiable transactions. The contributions of this paper can be summarized as follows:

(1) An integrated scheduling and trading (ISAT) framework is introduced that incorporates a cyber-physical blockchain system coupled with a smart contract-based V2G trading scheme. This framework is designed to enhance trust, transparency, and fairness in the energy transaction within distributed power systems through the synergistic integration of scheduling and trading.

(2) A cyber-physical blockchain structure is developed, tailored to the insights of the distributed power system. This structure capitalizes on the dual capabilities of SCPs, which combines traditional metering functions with computing, communication, and storage within the blockchain network. Additionally, a fast consensus mechanism is employed to expedite transaction validation in adaptation to the localized nature of distributed power grids and balancing robustness with the requisite scalability for energy transactions.

(3) A V2G trading scheme is delineated underpinned by a game-theoretical pricing strategy aimed at maximizing profits for both the system operator and EVs. Powered by the usage of smart contracts, this strategy facilitates autonomous V2G decision-making by EVs, obviating the need for comprehensive market participant information, thereby simplifying the trading process, enhancing market efficiency, and preventing privacy leakage of EVs.

(4) The efficacy and applicability of the proposed ISAT framework are rigorously validated through a case study employing real-world data from the Southern University of Science and Technology (SUSTech). The evaluation results highlight the framework's potential scalability and robustness of blockchain consensus, and the case study attests to the framework's effectiveness of the proposed V2G scheduling method, making ISAT framework a compelling case for its broader application across various electricity market stakeholders.

The overall workflow of this research is given in Fig. 1, and the remainder of this article is organized as follows: Section 2 presents a literature review. Section 3 details the proposed cyber-physical blockchain architecture. Section 4 outlines the V2G trading scheme. Section 5 provides an evaluation of blockchain and offers case studies of V2G applications. Finally, Section 6 concludes the article.

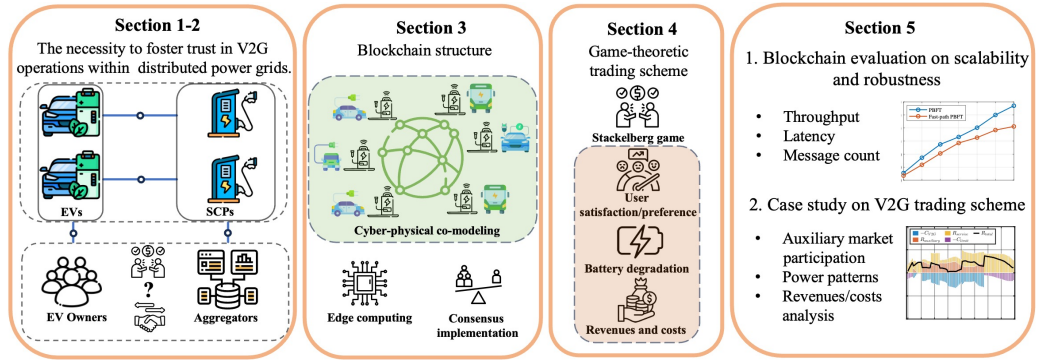

**Fig. 1.** Overall flow chart of this article.

## 2. Literature review

This section provides an analysis of the existing literature on V2G market participation, coordination and energy blockchain. It highlights the scope and depth of previous studies and identifies the research gaps that this paper seeks to fill. To systematically present the contributions of the examined studies, a summarized view is provided in Table 1.

### 2.1 V2G market participation and coordination

Many researchers have realized that the successful application of V2G requires its integration into business models and energy markets that ensure benefits for all stakeholders, particularly the EV owners [21], [22], [23]. In a distributed grid context, three key entities are involved: DSOs, EVAs, and EV users. For the DSO, it may participate in day-ahead bidding in the energy market or in the carbon market to maximize marginal revenue from V2G operations [24]. For instance, in [25], the focus is on how the DSO can leverage V2G to optimize its financial outcomes by participating in both the electricity market and the carbon emission market. The role of the EVA involves amassing the EVs under its control, setting scheduling objectives such as minimizing its own operational costs and tapping into auxiliary services markets like peak-valley energy trading, as detailed in [26]. The EVA essentially acts as a mediator between the grid and the EV users, coordinating the optimal scheduling of charging and discharging activities with typical approaches such as optimization or game theoretical modeling. For the EV user, participation in V2G operations can yield compensation through the EVA, such as by receiving payments for discharging back to the grid at high-demand times or by reducing personal costs through preferential pricing strategies for charging during low-demand periods [27]. Moreover, a vehicle-to-vehicle (V2V) market is proposed as a complementary to the V2G market when EV fully penetrates distributed power systems via a non-cooperative game [28].

V2G highly relies on effective coordination to integrate a growing fleet of plugged-in electric vehicles (PEVs) into the existing power grid. This coordination is primarily achieved through scheduling, broadly categorized into centralized and decentralized approaches [29]. Centralized scheduling encompasses the direct control of energy dispatch by PEV aggregators, taking into account the operational status of the grid and PEVs. Centralized scheduling can be further solved in a centralized approach, where a single optimization problem is solved for the entire system [30], or in a distributed approach, where the problem is broken down into smaller, localized sub-problems that are solved independently but coordinated through a central entity [31]. For instance, In Ref. [32], a centralized V2G optimization framework is developed to enhance the operational efficiency of airport microgrids by systematically coordinating EV charging and discharging schedules in alignment with flight operations. Leveraging distributed computing resources, Ref. [29] decreases the computational complexity by decomposing the overall scheduling problem into manageable sub-problems and builds up a load peak-shaving and valley-filling scheduling algorithm for V2G.

On the other hand, decentralized scheduling employs an incentive-based strategy where the operation behavior of PEVs is indirectly coordinated through dynamic pricing. Decentralized scheduling typically necessitates that grid operators or PEV aggregators disseminate pricing information to individual PEV users, and then each PEV user responds to price signals based on their personal preferences and objectives, thus fulfilling goals such as the redistribution of PEV charging loads from peak to off-peak periods [33]. The schemes of centralized and decentralized coordination are compared in Fig. 2. Ref. [34] proposed a decentralized energy consumption scheduling scheme, focusing on the integration of privacy preservation and aggregate load control within a day-ahead price-based demand side management program for a power distribution network. Ref. [35] examined decentralized V2G operation through a novel management concept for urban charging hubs, optimizing electric vehicle allocations and charging schedules to align with dynamic pricing signals and aggregator constraints, demonstrating improved grid stability and reduced charging demand unfulfillment.

## 2.2 Energy blockchain

The lack of a trusted intermediary presents substantial challenges in ensuring trust among V2G participants, establishing energy trade prices, and fulfilling agreements, whether automatically or imperatively. Blockchain technology, characterized by its security, transparency, immutability, and decentralization, emerges as a promising solution to enhance trust, transparency, and automation within decentralized V2G coordination frameworks. For example, In Ref. [36], a consortium blockchain-based energy trading scheme was introduced to facilitate demand response management in V2G systems, employing a peer-to-peer approach to optimize energy transactions between electric vehicles and service providers without requiring a trusted third party, thus improving transaction security and efficiency while ensuring privacy and data integrity. Ref. [37] presented a decentralized and privacy-preserving exchange scheme for V2G interactions with a fair exchange smart contract based on the hash-chain micropayment mechanism and a privacy-preserving protocol for V2G transactions. In Ref. [38], an energy blockchain environment was proposed, leveraging a game-theoretic approach to enhance trading efficiency based on the differentiation of market types based on supply and demand together with dynamic pricing models for both buyers' and sellers' markets. In Ref. [39], a blockchain-enabled, secure, fully decentralized energy management approach for multi-energy systems was developed with direct peer communication and an asynchronous updating mechanism to handle potential communication delays or interruptions. In Ref. [40], a decentralized V2G market trading system was developed using a hierarchical blockchain architecture, incorporating smart contracts for energy transactions that facilitate efficient and secure energy trading between electric vehicles, aggregators and the grid. These advancements underline the necessity for integrating IoT devices with blockchain to maintain real-time data accuracy, developing scalable blockchain architectures capable of processing large data volumes, and ensuring reliable synchronization between decentralized network operations and physical energy systems [41].

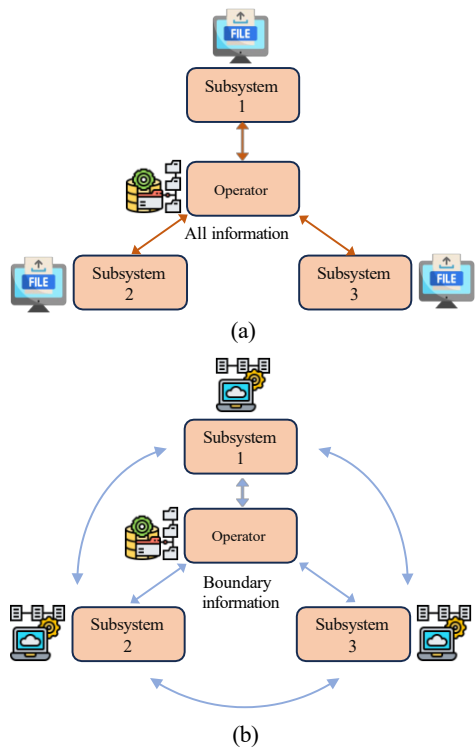

**Fig. 2.** Examples of different energy coordination schemes: (a) centralized coordination, and (b) decentralized coordination with blockchain implementation.

## 2.3 Research gap

While significant advancements have been made in V2G market participation, scheduling, and energy blockchain, there remains a gap between these developments and the large-scale implementation of V2G in real-world scenarios, encompassing a range of technical, economic, and social challenges [42], [43], [44]. One key issue that persists is the lack of a fair pricing mechanism and transparent trading processes for EV owners. Addressing this issue is critical for large-scale EV user participation while enabling the practical application of V2G systems in a way that stakeholders can benefit equitably.

The synthesis of blockchain into V2G systems leads to an open question, particularly in designing cyber structures that effectively complement the physical deployment and operations of energy resources. Notably, the physical elements can also bolster the cyber components by aiding critical functions such as identity validation through the actual physical connections, thereby enhancing the security and reliability of the trading process. Though significant research has been dedicated to centralized and decentralized scheduling within the V2G domain as well as energy blockchain, there is still a lack of detailed delineation of real-world scheduling and trading processes with the cyber-physical dynamics required for V2G systems. In particular, current studies often overlook the importance of establishing trustworthy environment for pricing, clearing, and settlement processes, which are essential for V2G operations in large-scale deployment in the context of distributed grids, involving EV users, EVAs, and DSOs. To bridge the gap toward scalable, real-world V2G implementation., this work proposes ISAT framework, where a cyber-physical blockchain structure is constructed considering the usage of SCPs with fast-path PBFT consensus, and a decentralized scheduling method is considered based on the Stackelberg game model to ensure fair outcomes between EV users and EVA.

**Table 1.** Comparisons of the proposed ISAT framework with other works

| Literature | Scheduling method | | Auxiliary service provision | Real-time control | Automation of trading | Cyber-physical co-modeling | Privacy-preserving | Blockchain simulation |
|---|---|---|---|---|---|---|---|---|
| | Centralized | Decentralized | | | | | | |
| [29] | ✓ | - | ✓ | ✓ | - | ✓ | ✓ | - |
| [32] | ✓ | - | ✓ | - | - | - | - | - |
| [34] | - | ✓ | ✓ | - | - | - | ✓ | - |
| [35] | - | ✓ | ✓ | ✓ | - | - | - | - |
| [36] | - | ✓ | ✓ | ✓ | ✓ | - | ✓ | ✓ |
| [37] | - | - | - | ✓ | ✓ | - | ✓ | ✓ |
| [38] | - | ✓ | ✓ | ✓ | ✓ | - | ✓ | - |
| [39] | - | ✓ | - | - | - | - | ✓ | - |
| [40] | - | ✓ | ✓ | ✓ | ✓ | - | ✓ | ✓ |
| [45] | ✓ | - | ✓ | ✓ | - | - | - | - |
| This work | - | ✓ | ✓ | ✓ | ✓ | ✓ | ✓ | ✓ |

## 3. Cyber-physical blockchain structure

The cyber-physical blockchain system studied in this paper is shown in Fig. 3. The system is mainly composed of the following entities.(1) Distributed power grid: The power grid architecture is distributed and integrates multiple generation sources in this system. It is engineered to manage both unidirectional and bidirectional flows of electricity, thus facilitating contributions from energy-prosumer entities, such as EVs, while simultaneously ensuring a consistent supply to standard consumers, including residential households and public facilities.

(2) EVs: In this model, EVs are considered to be prosumers, entities that both produce and consume energy. They are equipped with storage capacities and can inject energy back into the grid or draw energy as needed. The decision-making processes of these prosumers are influenced by economic incentives, as they adapt their energy dispatch in response to dynamic pricing and demand signals. Additionally, EVs will attend a blockchain-based scheduling and trading system when plugged into SCPs, which automatically schedules the charging and discharging activities and trades the energy on behalf of EVs.

(3) Aggregator: Aggregator acts as intermediaries between individual EV owners and the power grid. It aggregates the energy capacities of a fleet of EVs to sell or buy energy in bulk, enhancing the flexibility and reliability of energy management and making profits. Aggregator is not only instrumental in managing contracts and redistributing payments to EV owners but also in accruing profits through V2G administration fees. Concurrently, the aggregator is in negotiations with the upper-level system operator such as DSOs to secure incentives for fulfilling demand response obligations. Additionally, the aggregator utilized a blockchain-based trading system, operationalized through SCPs, ensuring transparency and security in energy transactions.

After outlining the key entities involved in V2G operations, sections 3.1 and 3.2 provide a detailed explanation of how blockchain can be integrated into the existing V2G infrastructure. Specifically, these sections elaborate on the customized integration of the cyber-physical blockchain structure within the V2G ecosystem, emphasizing its design to meet the demands of V2G operation, serving as a backbone for decentralized, secure, and automated EV scheduling and energy trading in distributed power grids.

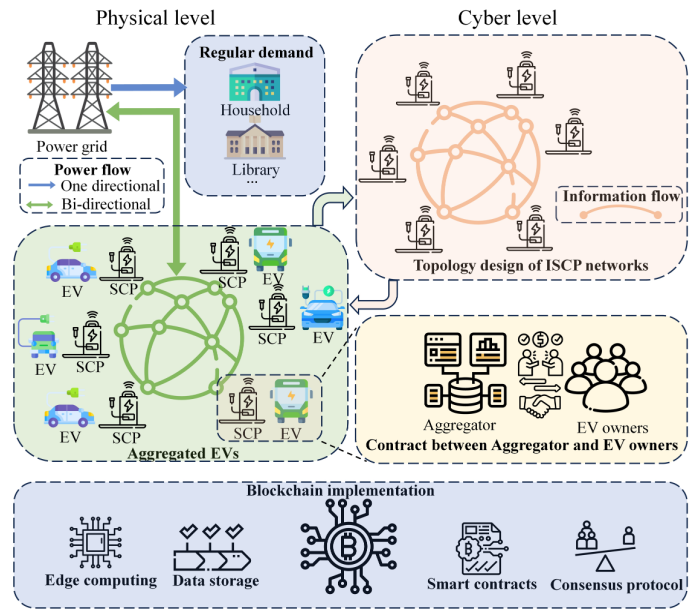

**Fig. 3.** Architecture of the proposed cyber physical blockchain structure.

## 3.1 Physical level: from distribution grid to smart charging points

Considering a regional V2G network orchestrated by an EVA, the EVs are denoted by $v_i$, $i=1,2,\cdots,N$, while the SCPs are denoted by $SCP_i$, $i=1,2,\cdots,M$, at the same time, the EV that is plugged in a certain SCP is denoted as $v_i^{plugged}$ and the corresponding SCP is denoted as $SCP_i^{plugged}$ to indicate its plugged-in state. When PEVs participate in charging activities within a region, they contribute to an increase in the electrical load. On the other side, the implementation of V2G allows PEVs to discharge their stored electricity back into the grid. This discharged electricity can be locally utilized for other purposes, such as demand response and other resilience enhancement measures [46], effectively acting as a mechanism to mitigate the regional electrical load [47]. As a form of encouragement, PEVs that partake in discharging activities receive compensation from the grid, which is determined by the discharging electricity price and the quantity of electricity discharged. Moreover, an additional operational mode is included where an EV can opt to remain "idle." In this state, the EV refrains from both charging and discharging activities, instead waiting for a subsequent period to reevaluate its decision. For this study, two primary assumptions are made: the charging/discharging prices are uniform across all PEVs within the distributed power grid and the electricity demands within a region are consolidated, forming an aggregated load on the grid. SCPs are at the core of this architecture, serving as energy control hubs. From a circuit topology perspective, each SCP comprises a three-phase full-bridge AC/DC converter, a DC-link capacitor, and a bidirectional buck-boost DC/DC converter. These components are crucial for efficient energy conversion and management. The SCPs connect to the utility grid through a passive filter at the front end, facilitating smooth energy flow. On the other hand, SCPs interact with PEVs, managing charging and discharging activities and acting as intermediaries between users and EVAs, providing a platform for users to engage in energy transactions.

## 3.2 Cyber level: from smart charging points to permissioned blockchain node

At the cyber level, SCPs are linked to form a decentralized communication network to facilitate the PEV charging and discharging. In the downlink, the price messages are broadcast to SCPs so that the PEVs can make decisions. In the uplink, the vehicular decision data are uploaded accordingly. Besides, SCPs collect vehicle information, control/measure average charging/discharging power over certain intervals and execute the smart contracts between users and EVA based on their computational capabilities. In all, this internet of SCPs functions as the backbone for the blockchain's communication and execution. In assessing network performance, the latency experienced by a message $M^{s\rightarrow r}$ traversing from a sender $s$ to a receiver $r$ is calculated by the following formula [48]:

$$Delay(M^{s\rightarrow r}) = L(s\rightarrow r) + T(M^{s\rightarrow r}) + P_r(M^{s\rightarrow r}) + Q_r(M^{s\rightarrow r}) \quad (1)$$

where $L(s\rightarrow r)$ quantifies the propagation delay as the signal traverses the medium separating $s$ and $r$, which is often calculated by the interconnecting distance and signal's propagation speed. $T(M^{s\rightarrow r})$ is the representative of the transmission delay, which is the quotient of the message size and the minimum bandwidth between $s$ and $r$, reflecting the temporal extent required to dispatch the message at the sender's transmission rate. The variables $P_r$ and $Q_r$ articulate the processing and queuing delays at the receiver $r$, respectively. Processing delay encompasses the duration for the receiver to handle the message, which may involve error checking and prioritization, while queuing delay accounts for the time the message lingers in the queue prior to processing.

In the process of data routing, such as uploading transactions to the blockchain, achieving consensus is essential for nodes to validate information so that information can be recorded in a transparent and immutable way. The widely recognized PBFT consensus protocol unfolds across three principal phases: pre-prepare, prepare, and commit. These stages are meticulously designed to secure a reliable and fault-tolerant consensus within distributed systems, as detailed by [49]. The protocol initiates with a selected leader node disseminating a PRE-PREPARE message to all peers, verifying the request's authenticity and consistency. Upon receiving a valid PRE-PREPARE message, nodes transmit a PREPARE message, signifying their agreement to engage in the consensus process. Finally, nodes issue COMMIT messages, and consensus is deemed achieved once a supermajority (over two-thirds) of the nodes have dispatched their commit messages. While the PBFT protocol enhances the reliability and fault tolerance of blockchain transactions, it also imposes a considerable workload on the system. This increased demand stems from the protocol's rigorous consensus mechanism, requiring multiple communication stages among nodes to verify transactions. This complexity can lead to higher computational and bandwidth requirements, impacting the system's overall efficiency and scalability.

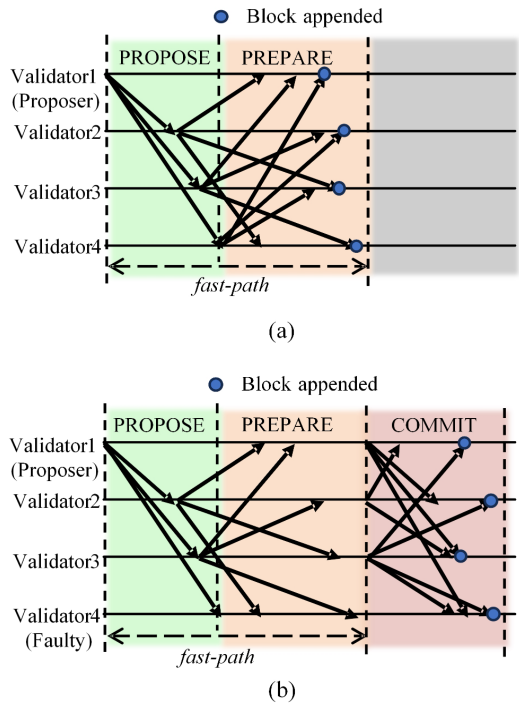

**Fig. 4.** Communication patterns in the fast-path PBFT consensus protocol with different states: (a) scenario within the fast-path timeframe, (b) scenario with a faulty validator 4, resulting in a commit phase. The blue dots represent the point at which each node decides to add the new block and integrate it into its blockchain sequence.

To ensure scalability, a permissioned blockchain structure is implemented where each SCP functions as a node within the blockchain. Given the geographical and network proximity of these nodes, a "fast path" approach is adopted in the consensus step with the following features [50]:

(1) Correct operation under byzantine fault, tolerating less than one-third of byzantine validators.
(2) Dynamic membership in the validator set: an $\alpha$ % of nodes are selected to partake in the consensus process as validators in each round.
(3) A latency of only two message delays for block decision post-network synchrony, assuming known maximum message latency and validators are all honest.

The integration of the "fast-path" mechanism within a cyber-physical system specifically designed for real-time energy trading in V2G networks forms the backbone of the proposed V2G trading scheme in Section 4. By leveraging the proximity of SCPs, the adaptation of the fast-path PBFT consensus protocol enables faster transaction validation, reduces latency, and thereby improves system scalability. Further simulations and analyses demonstrating this adaptation are presented in Section 5.1.

Procedures under the proposed consensus algorithm in a single round are listed as follows:

(1) Each SCP possesses a pair of cryptographic keys: one public key that acts as a unique address, ensuring that all transactions and actions can be traced back to the operator, and one private key that is used for digitally signing messages, thereby securing the integrity of communications and transactions within the network. At each time slot, the SCPs that are occupied with normal functionality are grouped, forming a set $SCP^{plugged}$
(2) At the consensus process onset, a leader SCP $l$ is randomly elected for the round by modulating the hash of the latest block with the number of blocks from the set of $SCP^{plugged}$. And the validator set $C(l)$ will be chosen by the leader $l$. This validator set $C(l)$ will cover $SCP^{plugged}$ as much as possible while randomly selecting other unoccupied SCP as validators to fill the requirement of $\alpha$ % nodes. This leader proposes a new block containing a batch of transactions, which are signed with their private key to confirm authenticity and multicast it to the validators, as shown in Fig. 4(a).
(3) Upon receiving a PROPOSE message, validators, the nodes within the validator set proceed to the prepare phase by multicasting a PREPARE message back to the network. These messages serve as an endorsement of the proposed block, with validators also verifying that the block extends the current blockchain correctly and coincides with the current round number.
(4) If a validator amasses a supermajority (at least 2/3 of validators) of PREPARE messages within the configured 'fast-path' timeout threshold, it will finalize the block as shown in Fig. 4(a). Failing this, as shown in Fig. 4(b), the validator enters the commit phase, where the validator multicasts a COMMIT message to express its firm commitment to the proposed block. This phase involves the validators' digital signatures as a form of consensus. If the commit phase is unsuccessful or if the round timer expires, indicating a round timeout, the protocol dictates a round change to transition to the subsequent round.

(5) Following the commit phase, a validator that receives a supermajority of COMMIT messages finalizes the block. This finalized block is then appended to the blockchain, and a DECISION message is broadcast across the network. This message includes both the block and a certificate composed of all the COMMIT messages, signifying the network's consensus and the block's definitive addition to the blockchain. During a successful consensus, non-validator nodes passively receive only the PREPARE and COMMIT messages.

This consensus algorithm ensures that most SCPs are synchronized across the network, maintaining a coherent and tamper-resistant ledger. It is also worth noting that the synergy between SCPs at the physical level and the corresponding communication network at the cyber level creates a robust and scalable blockchain system. This dual-level approach leverages the strengths of both domains to create a secure and resilient framework for energy transactions and executions within the V2G network.

Besides consensus, a significant challenge in energy blockchain is its inherent limitation in interfacing with the external world. For instance, in the context of Ethereum [51], the network can achieve consensus on the outcome of computational sequences. However, these computations are constrained to operate solely on data existing within the blockchain data that has been previously inputted by an Ethereum user into the virtual machine's memory. The network's capacity is limited to reaching consensus on the presence of this data on the blockchain, rather than on its validity. Consequently, the role of trusted entities, known as oracles, becomes crucial. They serve to authenticate external data, thereby bridging the gap between the blockchain and the external world.

As of now, the integration of external data into smart contracts, such as real-time electricity price information, necessitates the use of oracles. These oracles function as third-party services that validate data from various web services. Once verified, this data is then written into the blockchain through a specialized smart contract. This process allows smart contracts to access and utilize external data by calling upon the oracle contract. The use of oracles expands the functional capacity of smart contracts, enabling them to interact and respond to real-world data and events [52].

## 4. V2G trading scheme

This section outlines the V2G methodology for managing energy trading operations between PEV users and the EVA. To ensure fair interactions between these two parties, this study adopt a game-theoretical approach that accounts for both the PEV users' interests and the EVA's objectives in managing energy storage and discharging operations. Specifically, a Stackelberg game model is utilized, where the EVA acts as the leader and the PEV users as the followers [53]. This joint modeling approach helps to address the diverse needs of both parties while mitigating potential conflicts of interest. In addition, the use of smart contracts to automate agreements between PEV users and the EVA. These smart contracts streamline the transactional process, ensuring transparency, reliability, and automation in V2G trading operations, enhancing fairness and reducing transaction costs.

### 4.1 Game description

In the proposed V2G trading scheme, this research adopts a Stackelberg game model to represent the interaction between the EVA and the PEV users. The Stackelberg game is used to model hierarchical decision-making processes, where one party makes a move first, and the other parties react to this move. In a typical Stackelberg game, the leader makes decisions that influence the actions of other participants. The leader has the advantage of acting first, setting the terms such as prices, quantities, or policies that the followers will respond to. The followers observe the leader's move and then make their own decisions based on the leader's actions, optimizing their own objectives in response to the leader's strategy [54]. In the context of V2G trading, the EVA acts as a leader, controls key pricing decisions, and tries to maximize its revenue. PEV users act as followers, responding based on their individual utilities. The EVA, as the leader, anticipates or knows how the PEV users will react to its pricing strategy and adjust its actions accordingly to achieve optimal outcomes for itself while indirectly fostering mutual benefits [53]. The detailed modeling of the PEV utility function and EVA revenue function is discussed later in this section. A smart contract is subsequently implemented to automatically execute the Stackelberg game process within the cyber-physical blockchain structure.

## 4.2 PEV model

The utility function for a PEV user encapsulates three components: the satisfaction of charging or discharging, the cost associated with battery degradation, and the capital cost or gain through the charging and discharging activities. For PEV $v_i^{plugged}$ in the time slot $t$, let $SoC_i^t$ represent the state of charge (SoC) and $x_i^t$ denote the normalized quantity of electricity charged or discharged while $E_0$ acts as the standard quantity of the electricity. $E_i^t$ measures its current amount of electricity and $E_i^{capacity}$ measures the total capacity of electricity for $v_i^{plugged}$. Let $p_{real-time}^t$ and $L^t$ denote the real-time electricity price and the electricity load at time slot $t$, respectively, while $p_c^t > 0$ represents the charging price and $p_d^t < 0$ indicates the discharging price at the time slot $t$. A day is divided into 96 timeslots, with each time slot being 15 minutes.

### 4.2.1 User satisfaction

User satisfaction during both charging and discharging is an essential component of the utility function. This satisfaction is modelled to reflect how users perceive the value of the electricity charged or discharged, with different sensitivities for the two scenarios. A logarithmic function is used to account for diminishing returns in charging and heightened sensitivity in discharging.

The user satisfaction $S_i^t$ during charging is modelled as follows:

$$S_i^t = w_i^t ln(1+x_i^t), 0 \leq x_i^t \leq 1, \tag{2}$$

$$w_i^t = \frac{\beta_i}{SoC_i^t}, \tag{3}$$

where $w_i^t$ reflects the willingness to charge, with $\beta_i$ being a predefined constant reflecting the preference of user $i$. The willingness is inversely related to the remaining SoC, which indicates that when the $SoC_i^t$ is low, $v_i$ tends to have a higher willingness $w_i^t$ to charge. The logarithmic function $ln(1+x_i^t)$ captures the diminishing growth in satisfaction as more energy is charged. As $x_i^t$ increases, satisfaction grows, but at a slower rate, reflecting the fact that users place more value on the initial units of charged energy when their SoC is low. This diminishing return ensures that while charging more energy does increase satisfaction, the additional benefit becomes less significant as more energy is charged. Thus, the model aligns with the intuitive idea that users derive greater utility from the first few units of charged energy, and the marginal utility decreases as charging continues.

For discharging scenarios, where $x_i^t$ is assumed to be a negative value $(-1 \leq x_i^t < 0)$, the satisfaction function is adjusted to account for the heightened sensitivity to energy loss:

$$S_i^t = w_i^t (ln(2 + x_i^t) - 1), -1 \leq x_i^t < 0. \tag{4}$$

The function $ln(2 + x_i^t) - 1$ is designed to reflect that users are more sensitive to discharging, meaning their satisfaction decreases more rapidly as they discharge electricity. In contrast to charging, where satisfaction grows slowly, discharging leads to a sharper decline in satisfaction, indicating that users perceive the loss of energy as more significant. This reflects the fact that discharged electricity causes a drop in satisfaction, highlighting the increased sensitivity to energy loss compared to the relatively slower satisfaction growth experienced during charging.

The use of a logarithmic function in both charging and discharging aims to promote proportional fairness, as it models diminishing returns for charging and heightened sensitivity for discharging. This general satisfaction model is chosen because it effectively captures user behaviour in both scenarios. It balances fairness in energy allocation during charging while acknowledging users' aversion to energy loss during discharging, making it an intuitive while generalizable model for V2G operations [55].

### 4.2.2 Battery degradation cost in V2G interactions

Battery degradation in V2G systems encompasses two dimensions: calendar life and cycle life. Calendar life refers to capacity loss over time, a factor not directly related to battery usage but affected by environmental conditions where the battery is stored. Cycle life, however, is determined by the frequency and depth of battery charging and discharging cycles, reflecting operational wear and tear. In the utility function, a specific concentration is given to evaluate the costs related to cycle life degradation.

The cycle life of a battery, representing the maximum number of charge-discharge cycles it can undergo, is heavily influenced by the depth of discharge (DoD). The DoD refers to the percentage of the battery's total capacity that is used during each cycle. A deeper discharge (higher DoD) leads to more significant degradation, reducing the overall cycle life of the battery. This relationship is modeled using a best-fit curve based on empirical data, which captures how increased discharge depths exponentially reduce the number of available cycles [56]. Mathematically, this relationship can be expressed as:

$$N_{cycle} = \kappa_0 \cdot DoD^{-\kappa_1} \cdot e^{\kappa_2 (1-DoD)}, \tag{5}$$

where $\kappa_0$, $\kappa_1$ and $\kappa_2$ are coefficients derived from curve fitting, specific to the battery's characteristics and manufacturer data. This formula reflects the physical reality that deeper discharges cause more stress on the battery, accelerating wear and tear. As DoD approaches 1 (full discharge), the available cycle life significantly decreases due to this compounding effect of degradation.

The cost of battery degradation per cycle, $C_E(SoC)$ represents the financial cost associated with the wear on the battery from each charge-discharge cycle. This cost is influenced by DoD, as deeper discharges lead to greater degradation. The degradation cost is also proportional to the battery's capital cost, the cost related to the purchase of the battery. The formula for the degradation cost is:

$$C_E(SoC) = \frac{C_{cap}}{N_{cycle}} \tag{6}$$

where $C_{cap}$ represents the capital cost of the battery. Essentially, this equation calculates the cost of degradation as the per-cycle share of the battery's total cost, assuming that the battery's life is directly tied to the number of cycles it can perform before a significant capacity loss occurs.

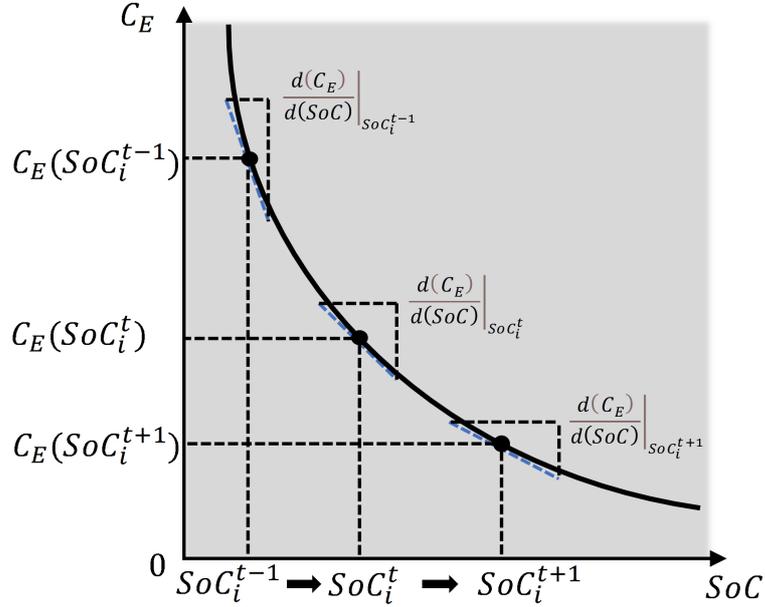

**Fig. 5.** One cycle cost function.

To simplify the computation of battery degradation costs, especially when considering real-time adjustments in the SoC, piecewise linearization is applied to the degradation cost function [57]. This technique involves approximating a non-linear function using linear segments for ease of calculation, particularly in optimization problems, as shown in Fig. 5. By using the first-order Taylor expansion, the degradation cost is approximated around the current SoC, making the calculations more computationally efficient for scheduling algorithms. Mathematically, this can be expressed as:

$$C_E(SoC_i^t) \approx C_E(SoC_i^{t-1}) + \frac{d(C_E)}{d(SoC)}\Big|_{SoC_i^{t-1}} \cdot (SoC_i^t - SoC_i^{t-1}) \tag{7}$$

The battery degradation cost $BDC_i^t$ during charging or discharging is calculated by assessing the incremental wear on the battery based on changes in its SoC. The degradation cost for a given time slot $t$ is influenced by how much the SoC changes during that time, reflecting the additional wear caused by each charging or discharging event. This can be modelled as:

$$BDC_i^t = C_E(SoC_i^{t-1}) - C_E(SoC_i^t) \approx a_{i,SoC_i^{t-1}} \cdot |x_i^t|, \tag{8}$$

where $a_{i,SoC_i^{t-1}} > 0$ is a parameter that is non-linearly related to the $SoC_i^t$. and reflects the battery's degradation rate based on its current charge level. The change in degradation cost depends on the absolute change in the quantity of electricity $x_i^t$ charged or discharged. This model assumes that both charging and discharging contribute equally to cell degradation and that the impact of each cycle remains consistent across the battery's lifetime.

### 4.2.3 Final utility function

The final utility function for a PEV user, $u_i^t(x_i^t, p_c^t, p_d^t)$ integrates the satisfaction from charging or discharging, the battery degradation cost, and the capital cost or gain through the charging and discharging activities. It is given by:

$$u_i^t\left(x_i^t, p_c^t, p_d^t\right) = \begin{cases} w_i^t \ln\left(1 + x_i^t\right) - p_c^t x_i^t, & 0 \le x_i^t \le 1 \\ w_i^t \left(\ln\left(2 + x_i^t\right) - 1\right) - 2BDC_i^t + p_{d,i}^t x_i^t, & -1 \le x_i^t < 0 \end{cases} \tag{9}$$

For charging $(0 \le x_i^t \le 1)$, the utility reflects the satisfaction from charging minus the cost of energy. For discharging, the utility includes the satisfaction from discharging, the battery degradation cost, and the compensation from discharging. Noted that to account for battery wear, particularly during discharging, which is often initiated by the EVA and necessitating a recharge, $-2BDC_i^t$ is incorporated into the utility function in discharging condition.

### 4.2.4 Critical charging/discharging price

Under varying charging and discharging prices, PEV can make decisions to charge or discharge, prioritizing its self-interest. A game-theoretical method is employed to analyze the conditions under which a PEV opts for charging or discharging. First, in the case of charging at a certain charging price $p_c^t$, the utility function's derivative with respect to unit charging quantity $x_i^t$ is given by:

$$\frac{\partial u_i^t\left(x_i^t, p_c^t\right)}{\partial x_i^t} = \frac{w_i^t}{\left(1 + x_i^t\right)} - p_c^t, \tag{10}$$

$$\frac{\partial^2 u_i^t\left(x_i^t, p_c^t\right)}{\partial x_i^{t^2}} = -\frac{w_i^t}{\left(1 + x_i^t\right)^2} < 0, \tag{11}$$

which indicates the concavity of the charging utility function. Setting $\frac{\partial u_i^t\left(x_i^t, p_c^t\right)}{\partial x_i^t} = 0$ and solving for $x_i^t$ yields the optimal response:

$$x_i^{t*} = \frac{w_i^t}{p_c^t} - 1, \tag{12}$$

where $x_i^{t*}$ is considered as the best response of $v_i^{plugged}$ at time slot $t$, which maximizes the utility on the condition of charging price $p_c^t$. The best utility function can be calculated by substituting $x_i^{t*}$ into charging utility $u_i^t(x_i^t, p_c^t)$ as:

$$u_i^{t^*}\left(p_c^t\right) = w_i^t \ln\left(\frac{w_i^t}{p_c^t}\right) - w_i^t + p_c^t, \tag{13}$$

which can be verified as a convex function while being non-negative on the condition that $p_c^t, w_i^t$ is positive. Letting $x_i^{t^*} = 0$, the critical charging price (CCP) $CCP_i^t$ of $v_i^{plugged}$ can be determined as:

$$CCP_i^t = w_i^t \tag{14}$$

CCP acts as a decision-making threshold for charging or discharging. If $0 \le p_c^t < CCP_i^t$ the best utility can be attained with $x_i^{t^*} > 0$ and accordingly $u_i^{t^*} > 0$. In this circumstance, the $v_i^{plugged}$ is willing to charge. Otherwise, $v_i^{plugged}$ is not expected to charge.

Now considering the case of discharging, for the discharging utility in (9), it is found that:

$$\frac{\partial u_i^t\left(x_i^t, p_d^t\right)}{\partial x_i^t} = \frac{w_i^t}{\left(2 + x_i^t\right)} + p_{d,i}^t + 2a_{i,SOC_i^{t-1}}, \tag{15}$$

$$\frac{\partial^2 u_i^t\left(x_i^t, p_d^t\right)}{\partial x_i^{t^2}} = -\frac{w_i^t}{\left(2 + x_i^t\right)^2} < 0, \tag{16}$$

which indicates the concavity of the discharging utility function with respect to $x_i^t$. Setting $\frac{\partial u_i^t(x_i^t, p_d^t)}{\partial x_i^t} = 0$, the best response is given by:

$$x_i^{t^*} = -\left(2 + \frac{w_i^t}{p_d^t + 2a_{i,SoC_i^{t-1}}}\right), \tag{17}$$

where $x_i^{t^*}$ maximizes the utility on the condition of discharging price $p_{d,i}^t$ at time slot $t$. And the best utility function can be calculated by substituting $x_i^{t^*}$ into charging utility $u_i^t(x_i^t, p_d^t)$ as:

$$u_i^{t^*}\left(p_d^t\right) = w_i^t\left(\ln\left(-\frac{w_i^t}{p_d^t + 2a_{i,SoC_i^{t-1}}}\right) - 2\right) - 2\left(p_d^t + 2a_{i,SoC_i^{t-1}}\right), \tag{18}$$

which can be verified as a convex function on the condition that $p_d^t, p_d^t + 2a_{i,SoC_i^{t-1}}$ are negative and $w_i^t, a_{i,SoC_i^{t-1}}$ are positive. Letting $x_i^{t^*} = 0$ the critical discharging price (CDP) $CDP_i^t$ of $v_i^{plugged}$ can be determined as:

$$CDP_i^t = -\frac{w_i^t}{2} - 2a_{i,SoC_i^{t-1}} \tag{19}$$

If $p_d^t < CDP_i^t \le 0$, the optimal utility is achieved when $x_i^{t^*} < 0$ and $u_i^{t^*} > 0$. In this case, the $v_i^{plugged}$ would like to discharge. Otherwise, $v_i^{plugged}$ is not expected to discharge.

### 4.2.5 PEV decision-making strategy

The best response function $x_i^{t^*}(p_c^t, p_d^t)$ for the charging, discharging and idle mode of V2G interactions can be summarized as follows:

$$x_i^{t^*}\left(p_c^t, p_d^t\right) = \begin{cases} \dfrac{w_i^t}{p_c^t} - 1, & \text{when } v_i^{plugged} \text{ is charging,} \\ -\left(2 + \dfrac{w_i^t}{p_d^t + 2a_{i,SoC_i^{t-1}}}\right), & \text{when } v_i^{plugged} \text{ is discharging,} \\ 0, & \text{when } v_i^{plugged} \text{ is idle.} \end{cases} \quad (20)$$

The decision-making rules for a PEV $v_i^{plugged}$ at time slot $t$ can be summarized as follows [58]:
(1) If the charging price is less than $CCP_i^t$ and the discharging price is greater than or equal to $CDP_i^t$, i.e., $CCP_i^t > p_c^t \geq 0 \geq p_d^t \geq CDP_i^t$, the PEV prefers to charge. In this case, $v_i^{plugged}$ at time slot $t$ will charge the battery with the corresponding best response amount of energy $x_i^{t^*}$.
(2) If the charging price is greater than or equal to $CCP_i^t$ and the discharging price is less than $CDP_i^t$, i.e., $p_c^t \geq CCP_i^t \geq 0 \geq CDP_i^t > p_d^t$, $v_i^{plugged}$ at time slot $t$ prefers to discharge the battery with the corresponding best response amount of energy $x_i^{t^*}$.
(3) If both the charging price and the discharging price are greater than or equal to their respective critical prices, i.e., $p_c^t \geq CCP_i^t \geq 0 \geq p_d^t \geq CDP_i^t$, $v_i^{plugged}$ prefers to remain idle during this time slot.
(4) If the charging price is less than $CCP_i^t$ and the discharging price is also less than $CDP_i^t$, i.e., $CCP_i^t > p_c^t \geq 0 \geq CDP_i^t > p_d^t$, $v_i^{plugged}$ will evaluate the utilities of both charging and discharging and opt for the action that yields the higher utility, choosing best response amount of energy $x_i^{t^*}$ accordingly.

The rules outlined above and the corresponding PEV decisions are illustrated in Fig. 6.

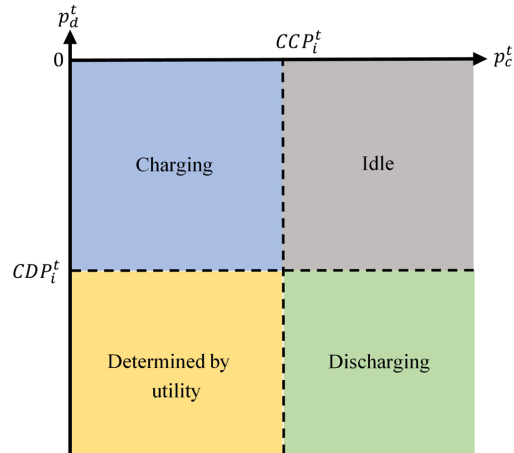

**Fig. 6.** PEV decisions under different charging and discharging prices.

## 4.3 EVA model

The EVA plays an integral role in V2G operations by strategically managing the discharging price $p_d^t$ thereby indirectly influencing the charging, discharging, and idle states of EVs, to meet the demands stipulated by the upper-level system operator and maximize its revenue. In the charging phase, the EVA charges EVs with dynamic, real-time electricity prices $p_{real-time}^t$ together with a certain service fee $W_{service}^t$. In the idle phase, the EVA incurs a nominal fee $p_{delay}^t < 0$ as the cost to pay for EVs' acceptance of this specific charging mode together with the delay. In the discharging phase, the EVA takes into account the auxiliary service demand $E_{limit}^t$ issued by the upper-level system operator, as well as the $CCP_i^t$ and the $CDP_i^t$ for each vehicle $v_i^{plugged}$ together with their best response function $x_i^{t*}(p_c^t, p_d^t)$ of the energy they will charge/discharge. Subsequently, the EVA adjusts the discharging price $p_d^t$ to optimize its revenue generation. The revenue maximization problem of EVA, denoted as $R_{total}$ is formulated as follows:

$$R_{total} = \sum_{t \in T} R^t \tag{21}$$

$$\max_{p_d^t} R^t = -C_{V2G}^t - C_{limit}^t + R_{grid, V2G}^t + R_{service}^t \tag{22}$$

where $C_{V2G}^t$ corresponds to the cost incurred by the aggregator for V2G services at time slot $t$, calculated as:

$$\begin{aligned} C_{V2G}^t &= \sum_{v_i^{plugged} \in V_{idle}^t} p_{delay}^t + \sum_{v_i^{plugged} \in V_{charge}^t} \left( p_c^t - p_{real-time}^t \right) E_0 x_i^{t*} + \sum_{v_i^{plugged} \in V_{discharge}^t} p_d^t E_0 x_i^{t*} \\ &= \sum_{v_i^{plugged} \in V_{idle}^t} p_{delay}^t + \sum_{v_i^{plugged} \in V_{discharge}^t} p_d^t E_0 x_i^{t*}, \end{aligned} \tag{23}$$

in which $V_{charge}^t$ represents the set of PEV $v_i^{plugged}$ at the state of charging during the time slot $t$, $V_{discharge}^t$ represents the set of PEV $v_i^{plugged}$ at the state of discharging during the time slot $t$, $V_{idle}^t$ represents the set of PEV $v_i^{plugged}$ at the state of idle during the time slot $t$. $p_c^t$ and $p_{real-time}^t$ cancel out since the charging price is set to be equal to the real-time electricity price in the retail market. For convenience, $V^t$ signifies the union of $V_{charge}^t$, $V_{discharge}^t$ and $V_{idle}^t$. $C_{limit}^t$ denotes the penalty for EVA as the result of not fulfilling the auxiliary service commitment at time slot $t$ and can be expressed as:

$$C_{limit}^t = \begin{cases} 0, & \text{if } \sum_{v_i^{plugged} \in V_{discharge}^t} E_0 |x_i^t| \geq E_{limit}^t, \\ \left( E_{limit}^t - \sum_{v_i^{plugged} \in V_{discharge}^t} E_0 |x_i^t| \right) \cdot \delta_t, & \text{otherwise,} \end{cases} \tag{24}$$

in which $E_{limit}^t \geq 0$ is the minimal auxiliary service amount assigned by the upper-level system operator for the current system operator, while $\delta_t \geq 0$ epitomizes the penalty factor for not achieving the minimal demand. $R_{grid, V2G}^t$ signifies the revenue accrued at the auxiliary market during time slot $t$:

$$R_{grid, V2G}^t = W_{grid}^t \sum_{v_i^{plugged} \in V_{discharge}^t} E_0 |x_i^t|, \tag{25}$$

in which $W_{grid}^t$ the auxiliary service price offered by the upper-level system operator. $R_{service}^t$ indicates the service fee collected for the charging service provided by EVA at time slot $t$:

$$R_{service}^t = \sum_{v_i^{plugged} \in V_{charge}^t} W_{service}^t E_0 x_i^t, \tag{26}$$

in which $W_{service}^t$ is the service fee charged by EVA for each charging vehicle. The constraints for the EVA model are:

$$0 \leq SoC_i^t \leq 1, \forall v_i^{plugged} \in V^t, \forall t \in T \tag{27}$$

$$x_i^t = x_i^{t^*}\left(p_c^t, p_d^t\right), \tag{28}$$

$$\left(SoC_i^t - SoC_i^{t-1}\right) \cdot E_i^{capacity} = E_0 \cdot x_i^t, \forall v_i^{plugged} \in V^t, \forall t \in T, \tag{29}$$

$$p_{d,min}^t \leq p_d^t \leq p_{d,max}^t \leq 0 \leq p_{delay}^t, \forall t \in T, \tag{30}$$

in which constraint (27) indicates the limitation of $SoC_i^t$ for the PEV $v_i^{plugged}$ attended in distributed energy management. Constraint (28) indicates that $v_i^{plugged}$ will decide its trading decision as well as the trading energy amount based on the given $p_c^t$ and $p_d^t$, as discussed in section 4.2. Constraint (29) expresses the relationship between $SoC_i^t$ and the energy traded amount, in which $E_i^{capacity}$ represents the energy capacity of $v_i^{plugged}$. Constraint (30) imposes limitations on the discharging price.

## 4.4 Smart contract procedure

In this system, each PEV $v_i^{plugged}$ is equipped with a digital wallet and functions as a light client [59], facilitating transactions within the blockchain network without storing the entire blockchain. This configuration allows PEVs to engage in blockchain transactions efficiently, verifying only essential information, such as block headers, without the necessity to store the entire blockchain. Each PEV is also endowed with a unique digital signature, enabling authenticated transactions and preventing repudiation. Each SCP, operating as a full node, maintains a complete copy of the blockchain, performs parallel computations and validates transactions and blocks while conducting operational trading on behalf of the PEV it is physically linked with. Oracles bridge the system with the external world, providing secure, reliable, and timely data inputs, such as real-time electricity prices $p_{real-time}^t$ in the retail market [52].

The smart contract process is articulated through the following steps:

(1) Initialization: An EV $v_i$ is plugged into the SCP $SCP_i$ with a preset preference $\beta_i$ determined by the EV owner. $SCP_i$ verifies the identity of the EV owner and the PEV $v_i^{plugged}$.

(2) Each $v_i^{plugged}$ conveys requisite information to its corresponding $SCP_i^{plugged}$, signed by the digital signature of the wallet owner. $SCP_i^{plugged}$ validates the status of $v_i^{plugged}$ and performs computations for the EV model, determining variables such as $CCP_i^t$, and $CDP_i^t$. It also computes the best response function $x_i^{t^*}\left(p_c^t, p_d^t\right)$. Subsequently, $SCP_i^{plugged}$ disseminates relevant data through its communication network.

(3) The system reads the current time slot $t$, and assimilates essential data such as the real-time charging price $p_{real-time}^t$, the permissible range of discharging price $\left[p_{d,min}^t, p_{d,max}^t\right]$ for the current time slot, the auxiliary service task $E_{limit}^t$, and auxiliary service price $W_{service}^t$ as communicated by the upper-level system operator via oracles.

(4) EVA calculates the optimal discharging price:

$$p_d^{t^*} = arg \max_{p_d^t \in \left[p_{d,min}^t, p_{d,max}^t\right]} R^t\left(p_d^t\right). \tag{31}$$

This computation is distributed among available functional SCPs.

(5) Once the optimal $p_d^{t^*}$ is decided, $p_d^{t^*}$ is broadcast in the network. Each $SCP_i^{plugged}$ then calculates its response to the current real-time charging price and the established discharging price $p_d^{t^*}$. Following this, $SCP_i^{plugged}$ conducts the power transmission $x_i^t$ with the connected $v_i^{plugged}$ and $v_i^{plugged}$ pays the corresponding fees from its wallet.

(6) With the completion of the operational cycle, the system enters a preparatory phase for the upcoming time slot. Here, each $v_i^{plugged}$, as a light client, updates its internal blockchain ledger with the latest related transaction information by accessing the connected $SCP_i^{plugged}$ and storing it in PEV's internal memory, ready for the next operational cycle at step 2 [60].

## 5. Evaluation and case study
### 5.1 Blockchain evaluation

With consensus being one of the main factors affecting the performance of a blockchain system, understanding and evaluating the performance characteristics of protocols is critical. In this study, SymBChainSim [61], a novel simulation tool for dynamic and adaptive blockchain management, is adopted to evaluate the scalability and robustness of the blockchain framework amidst faulty nodes and in the face of network, workload, and variations over time. The efficacy of the consensus protocol is investigated in the proposed cyber-physical blockchain system, with its performance being assessed based on latency, throughput, and consensus message count metrics. These outcomes are compared to a state-of-the-art protocol, PBFT [49]. The evaluation pivots on three key metrics [62]:

(1) Latency: Latency refers to the average time elapsed from submitting a transaction to the network until its confirmation, measured in milliseconds.
(2) Throughput: Throughput is quantified as the average number of transactions a blockchain network can process within a given timeframe, typically expressed in transactions per second.
(3) Consensus protocol message count: Consensus protocol message count measures the average number of messages exchanged per node per second and serves as an indicator of the communication load within a blockchain network's consensus process.

In blockchain evaluation, the environment is configured with specific parameters to investigate the system's performance under the PBFT protocol and the customized fast-path PBFT protocol. It is important to note that in this study, each SCP can serve multiple EVs, thus the total number of nodes in simulations may appear lower than expected for a system involving thousands of EVs. Since each SCP aggregates the charging and discharging decisions of several EVs, the actual number of nodes, represented by SCPs in the blockchain network, remains manageable while maintaining scalability. This aggregation further enhances efficiency by allowing each SCP to handle a significant portion of the transactions within its locality. Each simulation is conducted over a duration of 10 minutes, with a transaction size set at 0.01MB. The geographical locations of nodes are determined by randomly selecting from the charging stations within the SUSTech campus. The queuing delay $Q_r$ and processing delay $P_r$ are set to be 0.01 milliseconds for each transaction, aiming to model realistic network conditions.

For the communication topology, each node is modeled to have eight neighbors and employs the gossip method for information exchange [63]. This entails that upon receiving a message, a node randomly selects a subset of its neighbors to forward the message to. These neighbors, in turn, repeat the process, ensuring the message propagates through the network. The bandwidth for each node is configured to average 5MB/s with a standard deviation of 0.5 MB/s, adhering to a Gaussian distribution to simulate real-world network variability. Message sizes are uniformly set at 0.002 MB. Transactions are generated at a speed that follows a Poisson process at an average of 1 request per second [62].

Regarding block generation and validation, the block creation time by the leader is designated as 0.1 milliseconds, and the block validation time by each validator is set as 0.5 milliseconds. Each node is configured to take 0.1 milliseconds to validate a message, with a timeout threshold for messages established at 20 milliseconds and a 'fast-path' timeout threshold set to 10 milliseconds. Without loss of generality, all nodes are assumed to be validators in the experiment. Visual representations employ red curves/bars to delineate the performance metrics of the proposed blockchain system, contrasting with blue curves/bars for the PBFT-based blockchain system.

### 5.1.1 Scalability evaluation

To evaluate the scalability of the proposed consensus algorithm, simulations are conducted across an expanding array of node configurations: 12, 14, 16, 18, 20, 22 and 24, in environment devoid of faulty or Byzantine nodes. In these simulations, three key metrics are measured: latency, throughput, and consensus message count, and compared with the baseline PBFT protocol. These simulations aim to assess how effectively the fast-path PBFT protocol scales as the number of nodes increases compared to the baseline.

As shown in Fig. 7(a), the fast-path PBFT consistently outperforms the baseline PBFT in terms of latency. This improvement is largely due to the integration of the fast-path mechanism, which reduces the number of communication steps required to reach consensus in normal conditions. While the increase in latency is hard to prevent as the number of nodes grows, the fast-path PBFT protocol maintains lower latency than PBFT, making it more suitable for environments with real-time energy trading demands in distributed power grids.

Fig. 7(b) presents the throughput results. The fast-path PBFT protocol sustains a higher transaction throughput compared to PBFT. However, as the number of nodes increases, a diminishing throughput trend is observed in both protocols. This decrease is attributed to the additional communication overhead introduced by the larger network, where more messages are required to achieve consensus. Despite this, the fast-path PBFT's ability to process more transactions per second relative to the PBFT demonstrates its scalability advantage.

As illustrated in Fig. 7(c), the message count metric offers insight into the communication overhead imposed by the consensus protocol. The fast-path PBFT generates fewer consensus messages than the baseline PBFT, thanks to its streamlined communication steps. It is observed that a decreasing trend in message count as the number of nodes grows, which can be attributed to the reduction in throughput as network size increases. This reduction in message count highlights the protocol's advantage in handling communication overhead to the node as network size expands.

Overall, the fast-path PBFT consensus algorithm exhibits multiple advantages when compared to the standard PBFT protocol in scalability evaluation. By reducing latency and communication overhead through the fast-path mechanism, the protocol is well-suited for V2G applications in distributed grids that demand real-time processing and efficient resource allocation, making it a promising candidate for use in larger distributed networks. Additionally, the aggregation of multiple EVs at each SCP enhances network manageability and scalability.

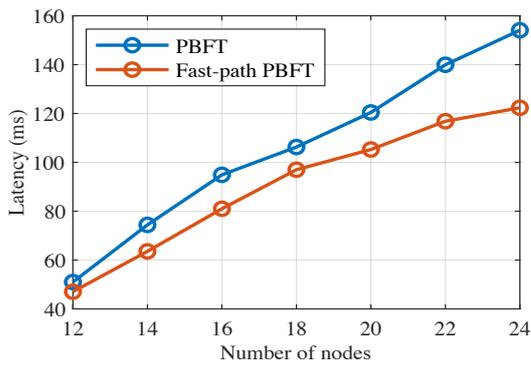

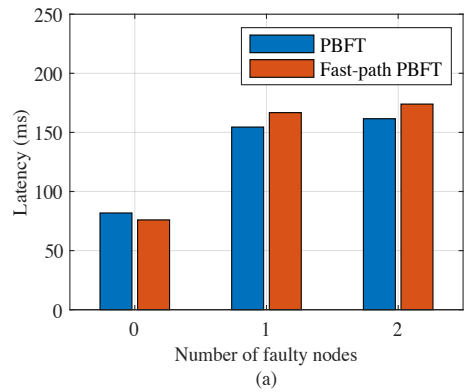

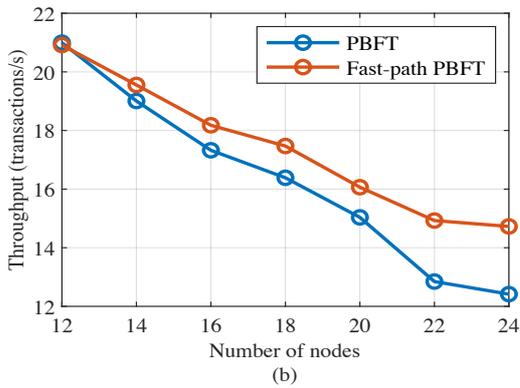

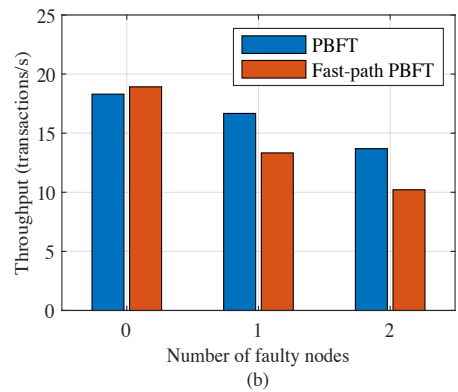

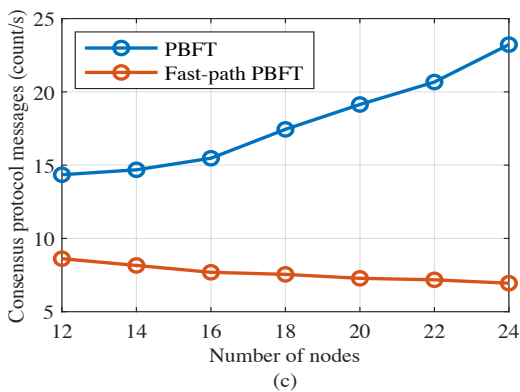

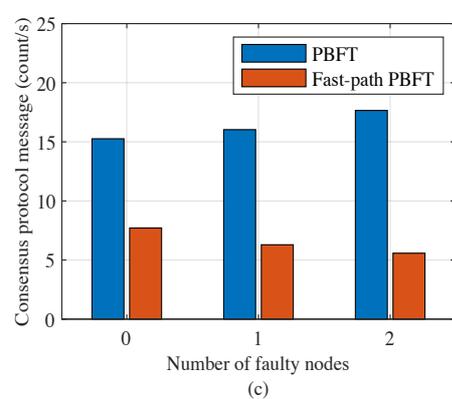

**Fig. 7.** Performance comparison between the fast-path PBFT protocol and the PBFT protocol as the number of nodes increases, (a) latency versus nodes, (b) throughput versus nodes, (c) Consensus messages court versus nodes.

**Fig. 8.** Performance comparison between the fast-path PBFT protocol and the PBFT protocol as the number of faulty nodes increases, (a) latency versus faulty nodes, (b) throughput versus faulty nodes, (c) consensus messages count versus faulty nodes.

### 5.1.2 Robustness evaluation

The robustness evaluation of the proposed consensus algorithm is conducted through simulations involving a fixed number of nodes of 15, under conditions incorporating 0, 1 and 2 faulty nodes. Faulty behavior is characterized by a fault occurrence randomly fluctuating between 30 milliseconds to 60 milliseconds in uniform distribution, implying an average fault onset ranging from 30 to 60 milliseconds into the simulation. Additionally, nodes are modeled to have a mean recovery time selected randomly between 3 milliseconds and 6 milliseconds in uniform distribution.

Fig. 8(a) reveals an increase in latency for both protocols as the number of faulty nodes rises. Notably, the fast-path PBFT protocol maintains a latency similar to that of the PBFT protocol under increasing fault conditions. At the same time, Fig. 8(b) illustrates a larger decrease in throughput for the fast-path PBFT protocol relative to the PBFT protocol with the increase of faulty nodes. This suggests a certain vulnerability in maintaining throughput under fault conditions in the fast-path PBFT protocol. A further aspect of analysis, depicted in Fig. 8(c), shows a decline in the consensus message count for the fast-path PBFT protocol. This reduction is attributed to the overall decrease in throughput, while the PBFT protocol experiences an increase in consensus message count, likely due to the intensified consensus process necessitated by faulty node scenarios.

In conclusion, the fast-path PBFT consensus protocol demonstrates commendable scalability, with a tolerable decrease in robustness compared to the PBFT protocol in scenarios involving faulty nodes. Despite this trade-off, it maintains comparable latency and a lower consensus message count, reflecting a balanced compromise between scalability and robustness when handling network faults. It is also important to note that these robustness concerns can be mitigated by incorporating procedural and error-checking mechanisms at both the hardware and software levels, including redundant communication paths, fault detection and recovery protocols, secure transmission design and tamper-resistant hardware components [64].

### 5.2 V2G trading scheme case study

The case studies are conducted using the historical EV routines obtained from SUSTech within the Guangdong electricity market in China. The study focuses on a diverse EV cohort of size 2,000, including electrical buses and privately owned EVs. The dataset encompasses EV parameters including battery capacity (in kWh), initial battery state of charge (in kWh), and the EV's arrival and departure times (in daytime), as detailed in Table 2 with 12 samples from the dataset. The pricing model for real-time charging electricity adopts the valley-peak pricing scheme implemented by the Shenzhen government. This model has three distinct pricing tiers: peak, valley, and flat rates, each applicable for specific periods within a day. The allocation of these rates is outlined in Table 3 [28]. While the dataset and pricing model are specific to Shenzhen, they mirror broader patterns commonly observed in urban electricity markets worldwide, where similar time-of-use (ToU) pricing schemes are employed to manage grid demand and encourage off-peak energy consumption. The peak-valley pricing structure in Shenzhen aligns with strategies used in many other cities that also implement TOU pricing models, thereby enhancing the relevance of this case study to global contexts with comparable mechanisms. Moreover, the driving behaviors and energy consumption patterns captured in SUSTech, a region with high EV penetration, provide valuable insights that extend beyond the local context, supporting the adaptability and broader applicability of the proposed V2G trading framework across diverse electricity markets [65].

**Table 2.** Driving patterns data of EVs (12 samples among 2000 records).

| EV sample IDs | 1 | 71 | 74 | 215 | 217 | 486 | 544 | 608 | 765 | 1509 | 1553 | 1766 |
|---|---|---|---|---|---|---|---|---|---|---|---|---|
| Battery volume (kWh) | 160 | 64 | 65 | 40 | 65 | 32 | 40 | 32 | 40 | 64 | 24 | 24 |
| Initial battery (kWh) | 57.2 | 19.9 | 17.7 | 26.2 | 37.1 | 20.2 | 20.6 | 13.4 | 25.9 | 47.6 | 21.4 | 9.8 |
| Arrival time (h) | 16:00 | 17:45 | 14:30 | 5:45 | 00:00 | 21:30 | 14:30 | 07:30 | 20:45 | 06:00 | 06:30 | 18:45 |
| Departure time (h) | next day 06:30 | next day 16:30 | next day 04:30 | 10:15 | 12:00 | next day 12:00 | next day 08:00 | 12:30 | next day 13:15 | 11:15 | 20:45 | next day 04:30 |

**Table 3.** Valley-peak prices (CNY/kWh) at each hour in Shenzhen.

| Time period | 00:00-08:00 | 08:00-10:00 | 10:00-12:00 | 12:00-14:00 | 14:00-19:00 | 19:00-24:00 |
|---|---|---|---|---|---|---|
| Type | Valley | Flat | Peak | Flat | Peak | Flat |
| Price | 0.26 | 0.66 | 1.12 | 0.66 | 1.12 | 0.66 |

In this case study, the existing SCPs in SUSTech campus are assumed to be upgraded to comply with proposed ISAT framework and are managed by an EVA. The DSO functions as the upper-level system operator for the SUSTech campus, issuing real-time auxiliary service demands and providing relevant information, such as real-time electricity prices from the market, through oracles. The DSO engages in day-ahead bidding within the Guangdong electricity market, making bids based on load forecasts for the following day [25]. Discrepancies between forecasted purchases $L_{pedicted}^t$ and actual consumption $L_{base}^t$ can result in either excess or deficit electricity procurement, resulting in penalties for the DSO. Fig. 9 showcases a comparison between the predicted and actual (or ideally forecasted) loads for the SUSTech campus on May 27 07:00 to May 28 07:00, 2021 based on the prediction model illustrated in [21]. The result shows that although DSO's model can capture the trend of future load, there still exists a certain amount of prediction error. This discrepancy indicates a deviation from the actual load behavior during that period and is highlighted by the red bars, indicating opportunities for V2G in the auxiliary services market to offset the DSO's potential penalties. For example, if the real-time imbalance exceeds 4% of the day-ahead bids, the DSO will receive a penalty along with a loss in credit from the Guangdong electricity market. To mitigate the risks of such financial penalties, the DSO leverages the flexibility provided by the EVA through the V2G framework in the scenario of positive deviation $L_{base}^t > L_{pedicted}^t$. In this scenario, the DSO will notify the EVA of the auxiliary service price $W_{grid}^t$ and the minimum required auxiliary service amount $E_{limit}^t$. This enables the DSO to align real-time load more accurately with the forecasted load by discharging energy from EVs when necessary, helping to reduce forecast errors and associated penalties.

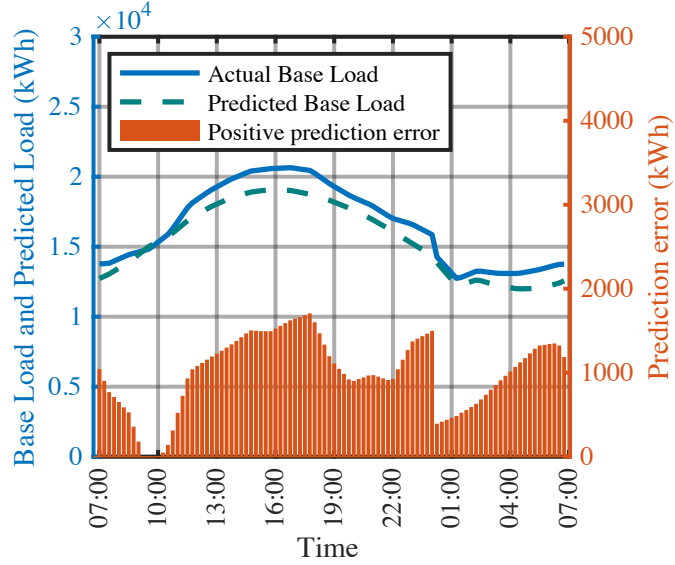

**Fig. 9.** The base load, predicted base load with positive prediction errors in SUSTech campus from May 27 07:00 to May 28 07:00 2021

When an EV owner plugs the vehicle, the SCP, now upgraded to comply with the ISAT framework, automatically identifies the vehicle and retrieves the owner's predefined information, such as the willingness to charge and the battery degradation parameters. Based on the owner's preferences and the real-time data, smart contracts deployed on the blockchain are automatically activated to handle the energy scheduling and trading between the EV and the EVA, as described in the smart contract process (Section 4.4). The decentralized computation capabilities of the SCPs allow for real-time scheduling. Each energy transaction between the EV and the grid is securely logged on the blockchain, ensuring data integrity, transparency, fairness, and automation. At the end of the charging or discharging session (e.g., when the owner returns to the vehicle), the smart contract is triggered to terminate the session and the final energy transaction is settled on the blockchain. The EV owner can review the session's details, including the amount of energy exchanged, the final cost, and revenue earned from providing auxiliary services to the grid. A copy of the blockchain blocks related to the owner's transactions is stored on the EV or another storage device for future reference. The parameters used in the V2G scheduling and trading simulation are detailed in Table 4 [26].

**Table 4.** Main parameters in the case studies

| Parameter | Description | Value |
|---|---|---|
| $W_{grid}^t$ | Auxiliary service price by upper-level system operator | 0.792CNY/kWh |
| $W_{service}^t$ | Service fee charged towards EV for charging service | 0.5CNY/kWh |
| $E_0$ | Maximum energy amount exchanged in 15 minutes | 3.75kWh |
| $E_{limit}^t$ | Minimal auxiliary service energy amount in positive deviation | $0.48\left(L_{base}^t - L_{predicted}^t\right)$ |
| $\delta_t$ | Penalty factor for not meeting $E_{limit}^t$ | $-3 \cdot p_{real-time}^t$ |
| $p_{delay}^t$ | Delay compensation for EV's idle state | -0.1 CNY/15min |
| $p_{d.max}^t$ | Upper bound for discharging price | 0 |
| $p_{d.min}^t$ | Lower bound for discharging price | $-3 \cdot p_{real-time}^t$ |
| $\beta_i$ | Preference constant for each individual PEV user | 0.81 |

Fig. 10 illustrates the bidirectional power flow from aggregated EVs participating in the proposed V2G scheme, along with the real-time electricity pricing and the calculated discharging price. Under the V2G framework, EVA aggregates information from the EVs to compute the optimal discharging price $p_d^t$ aimed at maximizing revenue and then broadcast it. The correlation between the discharging price and the real-time electricity price suggests that the EVA is able to adjust prices responsively in relation to market conditions. For example, during the period when the $p_{real-time}^t$ is at the "Peak" stage like 10:00 -12:00, EVA adjusts $p_d^t$ and captures a significant amount of discharging power with a relatively low cost.

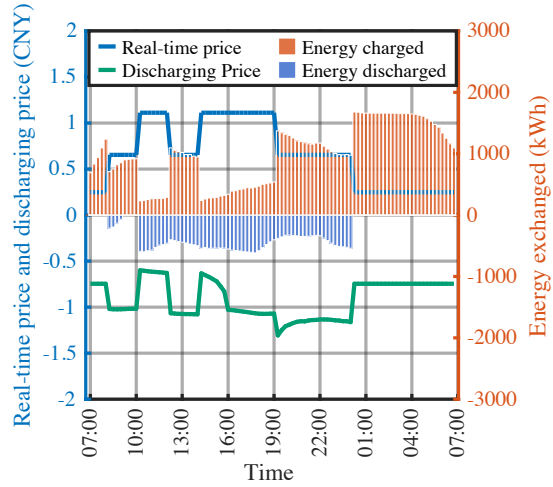

**Fig. 10.** Charging/discharging power with real-time electricity price and discharging price under the V2G scheme.

From the perspective of EV owners, the economic incentives presented in the V2G scheme are clear. PEVs react to $p_{real-time}^t$ and $p_d^t$, choosing to charge, discharge or remain idle based on their own strategies, as described in section 4.2. Notably, during hours when real-time electricity price $p_{real-time}^t$ surges, such as between 10:00 and 12:00, there is a discernible decrease in charging activity along with an increase in discharging activity. This indicates that EV owners are less inclined to charge and become comparatively more inclined to discharge. Conversely, the post-peak period, like 12:00 -14:00, sees a significant uptick in charging power as the electricity price returns to a 'flat' stage. During this time, the discharging power remains relatively stable, reflecting the increased readiness of EVA to offer higher incentives for discharging PEVs. Furthermore, the intensified charging activity observed during periods of the lowest electricity price underscores the cost-efficiency of the strategy adopted by PEVs.

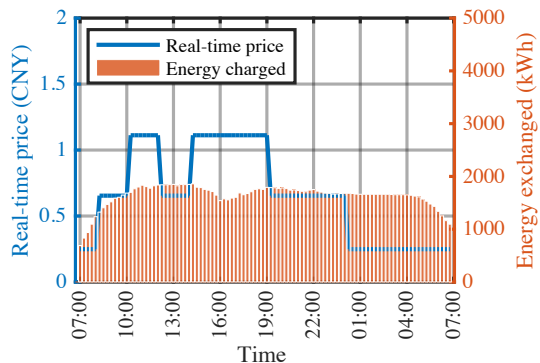

**Fig. 11.** Charging power with real-time electricity price under the UC scheme.

Fig. 11 demonstrates the charging behavior of aggregated EVs under an uncoordinated charging (UC) scheme devoid of V2G strategy. The depicted energy flow is unidirectional, with EVs conducting charging regardless of fluctuations in real-time electricity pricing. This figure highlights a rigid charging pattern, where EVs draw power from the grid even during high-price periods, such as between 10:00 and 12:00. Thus, without strategic interaction, the charging load here is less responsive to price signals, thereby demonstrating the potential benefits of V2G in managing demand and facilitating cost-effective energy consumption.

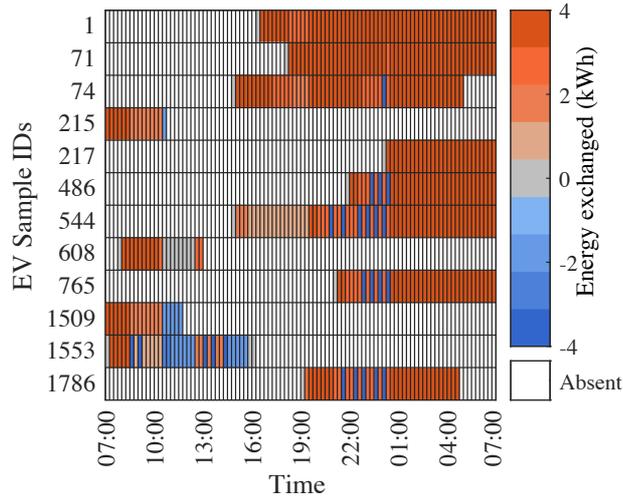

**Fig. 12.** Charging/discharging energy and plug-in duration of sample EVs.

Fig. 12 provides a detailed view of the energy exchange behaviors of sampled PEVs in Table 2 under the proposed V2G scheme. Each row represents an individual vehicle, with the color-coded bars indicating the magnitude and direction of power flow, with blue for discharging and red for charging, respectively. Absences are denoted by white spaces when vehicles are not plugged in. A strategic adaptation is demonstrated among the sampled PEVs, with each vehicle tailoring its energy exchange to accommodate its specific demand and availability in response to fluctuating temporal factors.

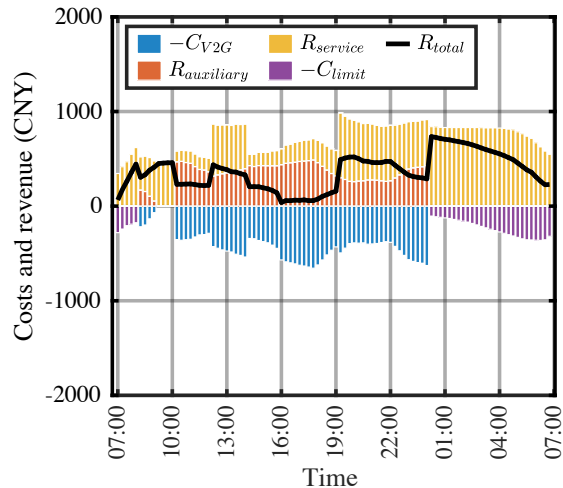

**Fig. 13.** Revenue and cost analysis of the V2G scheme.

Finally, Fig. 13 illustrates the financial dynamics, both costs and revenues, associated with the proposed V2G scheme, utilizing notations aligned with those introduced in Section 4. From the perspective of the EVA, this scheme allows for efficient utilization of PEVs to meet the minimum auxiliary service energy requirements assigned by DSO, thereby generating profits and reducing penalties. However, during periods when electricity prices are at their lowest, referred to as the 'Valley' stage, the EVA's ability to capitalize on the auxiliary services market diminishes. On the other hand, from the EV owner's standpoint, Table 5 offers a detailed comparison between the V2G and UC schemes. It highlights a significant 26% decrease in the average charging cost with an average discharging price 95% higher than the average charging price in absolute terms as compensation under the proposed V2G scheme. This strategic reduction in charging costs, combined with the energy contribution totaling -27, 346 kWh for V2G, showcases the PEVs' capacity to generate income through discharging activities, thus underscoring the economic effectiveness of the V2G approach.

Table 5. Comparisons between the V2G scheme and UC scheme in the case study

| Metrics | Total energy charged (kWh) | Total energy discharged (kWh) | Average charging price (CNY/kWh) | Average discharging price (CNY/kWh) |
|---|---|---|---|---|
| V2G | 95517 | -27346 | 0.4994 | -0.9717 |
| UC | 157485 | 0 | 0.6718 | 0 |

# 6. Conclusion

This study provides a technologically viable framework for the pivotal challenge of establishing trust, transparency, and fairness in vehicle-to-grid (V2G) energy exchanges, a key barrier hindering the large-scale adoption of V2G systems. Recognizing the lack of fair pricing mechanisms and transparent trading processes as a significant gap, a blockchain-enabled integrated scheduling and trading (ISAT) framework is proposed tailored to distributed power grids.

The ISAT framework introduces a cyber-physical blockchain architecture that leverages the dual functionalities of smart charging points (SCPs). By combining traditional metering with advanced computing, communication, and storage capabilities within the blockchain network, the framework enhances the integrity and security of energy transactions. The customized integration of a fast-path practical byzantine fault tolerance (PBFT) consensus mechanism accommodates the localized nature of distributed power grids, ensures rapid transaction validation, thereby supporting energy trading in the real-world scenario. Additionally, a game theoretical pricing strategy is considered to enable the dynamic and autonomous decision-making process for EVs and EV aggregators (EVAs), optimizing the overall system performance while maximizing mutual profits for both parties fairly. The streamlined smart contract process further enhances the automation and efficiency of the V2G process.

The efficacy of the ISAT framework was rigorously validated through a case study utilizing real-world data from the Southern University of Science and Technology (SUSTech). The evaluation demonstrated the framework's scalability, particularly in terms of latency, throughput, and message handling within the blockchain network with a tolerable decrease in robustness. The case study confirmed the effectiveness of the V2G scheduling method, showing a significant reduction in average charging costs for EV users compared to uncoordinated charging schemes and highlighting considerable economic benefits.

However, this study does not fully explore the wide-range of existing V2G scheduling methods that pursue diverse objectives. Future studies should delve into detailed comparisons of additional scheduling methods with different solving approaches and multiple objectives within the ISAT framework. Furthermore, investigating social behaviors related to EV charging and discharging, as well as assessing the long-term environmental implications of widespread V2G implementation, are important avenues for future research. As the energy market continues to evolve with emerging technologies and policies, ongoing research is imperative to adapt and refine the ISAT framework, ensuring its continued relevance and efficacy in the dynamic landscape of energy systems.

## CRediT authorship contribution statement

**Yunwang Chen:** Conceptualization, Investigation, Methodology, Validation, Formal analysis, Writing - Original Draft. **Xiang Lei:** Formal analysis, Writing - Review & Editing. **Songyan Niu:** Writing - Review & Editing. **Linni Jian:** Conceptualization, Supervision, Project administration.

## Declaration of competing interest

The authors declare that they have no known competing financial interests or personal relationships that could have appeared to influence the work reported in this paper.

## Acknowledgment

This work was supported by the Science and Technology Innovation Committee of Shenzhen under Project JCYJ20220530113008019.## Data availability

Data will be made available on request.

## References

[1] Mallapaty S. How China could be carbon neutral by mid-century. Nature 2020;586:482–3.
[2] International Energy Agency. Global EV Outlook 2021 2021.
[3] Williams G. Rare Earths Outlook 2019: EV Production to Drive Demand. CREI 2019;25:2–6.
[4] Jian L, Zhu X, Shao Z, Niu S, Chan CC. A scenario of vehicle-to-grid implementation and its double-layer optimal charging strategy for minimizing load variance within regional smart grids. Energy Conversion and Management 2014;78:508–17.
[5] Khalid MR, Alam MS, Sarwar A, Jamil Asghar MS. A Comprehensive review on electric vehicles charging infrastructures and their impacts on power-quality of the utility grid. eTransportation 2019;1:100006.
[6] Xu C, Behrens P, Gasper P, Smith K, Hu M, Tukker A, et al. Electric vehicle batteries alone could satisfy short-term grid storage demand by as early as 2030. Nat Commun 2023;14:119.
[7] Qin Y, Rao Y, Xu Z, Lin X, Cui K, Du J, et al. Toward flexibility of user side in China: Virtual power plant (VPP) and vehicle-to-grid (V2G) interaction. eTransportation 2023;18:100291.31